\documentclass[aps,pra,twocolumn,superscriptaddress]{revtex4-2}

\usepackage{physics}
\usepackage{siunitx}
\usepackage{amsmath}
\usepackage{gensymb}
\usepackage{graphicx}
\usepackage{dcolumn}
\usepackage{float}
\usepackage{wrapfig}

\begin{document}


\title{Lineshape Analysis of Double-Quantum Multidimensional Coherent Spectra}
\author{Bachana Lomsadze}
\email[]{blomsadze@scu.edu}
\affiliation{Department of Physics, Santa Clara University, Santa Clara, CA, 95050}
\affiliation{Department of Physics, University of Michigan, Ann Arbor, MI, 48109}


\author{Steven T. Cundiff}
\email[]{cundiff@umich.edu}

\affiliation{Department of Physics, University of Michigan, Ann Arbor, MI, 48109}



\date{\today}

\begin{abstract}
Double-quantum two-dimensional coherent spectroscopy (MDCS) is a powerful optical method that is used to study optical properties of atomic and complex molecular systems and semiconductor materials. Double-quantum 2D spectra and particularly the peak lineshapes on the spectra can also provide information about many-body interactions.  We model 2D spectra by solving the optical Bloch equations and show the effects of correlation between coupled resonances, which also explains the discrepancies between the experimental results reported by multiple groups. 
\end{abstract}


\maketitle

\section{Introduction}

\begin{figure*}
\includegraphics[width=1\textwidth]{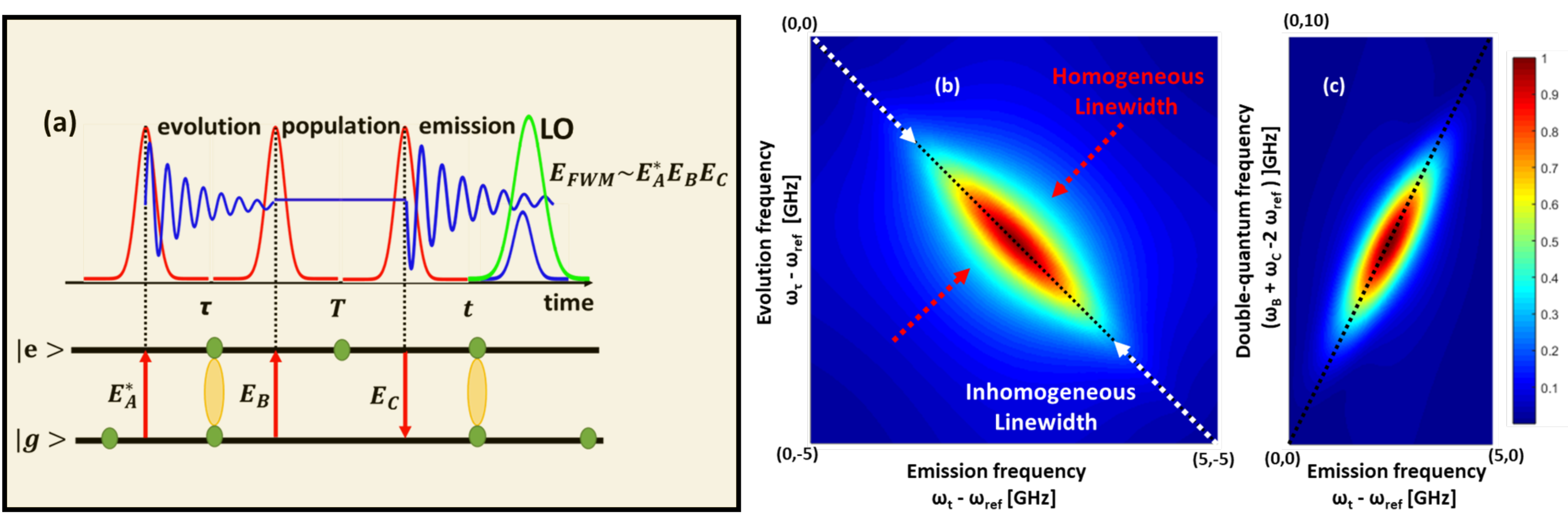}
\caption{\label{fig:MDCS_scheme} (a) Schematic diagram of multidimensional coherent spectroscopy. In the figure the photon echo excitation sequence is displayed. Pulse A* creates a coherence between the ground and excited states. Pulse B converts the coherence into the population state and then pulse C converts it back to the coherence between the ground and excited states which emits a Four-Wave-Mixing (FWM) signal. The signal is detected with a local oscillator pulse. $\ket{g}$ and $\ket{e}$ correspond to ground and excited states, respectively. (b)  Magnitude of a single quantum two-dimensional spectrum. The evolution frequency is negative to reflect the negative phase evolution during the evolution period in the photon echo excitation sequence. Black dotted line shows the diagonal line. White and red arrows indicate homogeneous and inhomogeneous linewidths.  (c) Magnitude of a double-quantum two-dimensional spectrum. The peak is elongated along the diagonal line (black dotted line).  $\omega_{ref}$ arbitrary optical frequency.}
\end{figure*}

In the past two decades multidimensional coherent spectroscopy (MDCS) \cite{stevemukamel, hamm2011concepts} has become a powerful and routine technique for studying optical properties and ultrafast dynamics of atomic/molecular samples and semiconductor materials \cite{LomsadzeScience, Thamer78, hamm2011concepts, Chris:review, LomsadzeNatPhot, Lomsadze2017, Tiwari:18, Lomsadzeieee}. MDCS is the only optical method that can simultaneously  measure homogeneous and inhomogeneously linewidth, identify coupling between the excited states, track energy redistribution in complex systems (in real time), and probe the many-body interactions. MDCS is based on concepts of Nuclear Magnetic Resonance experiments that is widely used for determining the molecular structure \cite{NMR}. A simplified schematic diagram of multidimensional coherent spectroscopy is shown in Fig.~\ref{fig:MDCS_scheme} (a). A sequence of three pulses (A, B, C) incident on the sample of interest generates a four-wave mixing (FWM) signal which is then heterodyne detected, using a local oscillator pulse, as a function of the delays between the excitation pulses. 
The recorded time domain interferogram is then Fourier transformed with respect to the time delays between the incident pulses and over the time period during which the signal is emitted to generate a multidimensional coherent spectrum.  

In MDCS, a multidimensional coherent spectrum generated by different pulse ordering provides different spectroscopic information. For example: One can use the photon echo excitation scheme \cite{photonecho} shown in Fig.~\ref{fig:MDCS_scheme} (a) (where the first pulse is the phase conjugated pulse (A*,B,C)) to measure homogeneous linewidth of inhomogeneoslsy broadened systems. In this excitation scheme pulse A* creates a coherence between the ground and excited states which then evolves in time. Pulse B converts the coherence into the population state and then pulse C converts it back to the coherence between the ground and excited states which emits a Four-Wave-Mixing (FWM) signal. A multidimensional spectrum is then generated by taking the Fourier transforms of the FWM signal with respect to $t$-emission and $\tau$-evolution times.  This spectrum is also referred to as a rephasing  single-quantum 2D spectrum (phase evolution during $t$-emission and $\tau$-evolution times are opposite and hence the resonances will recombine/rephase). A rephasing spectrum is shown in Fig.~\ref{fig:MDCS_scheme}(b) which  provides both the homogeneous and inhomogeneous linewidths of the sample and they can be extracted simultaneously \cite{Siemens:10}. In addition, lineshapes on a single quantum two-dimensional spectrum can also provide extremely valuable information. For example, one can investigate spectral diffusion (related to a correlation function  \cite{Hoffman13949, Ogilvie}) by measuring the elipticity of the elongated peaks (along the diagonal) on a single quantum two dimensional spectrum  as a function of the time delay between B and C excitation pulses \cite{Lazonder:06, spect:diff}. Over the years several methods have been developed to interpret lineshapes of single-quantum multidimensional coherent spectra \cite{Siemens:10, Richtereaar7697,Lazonder:06, spect:diff}. 

\begin{figure*}
\includegraphics[width=1\textwidth]{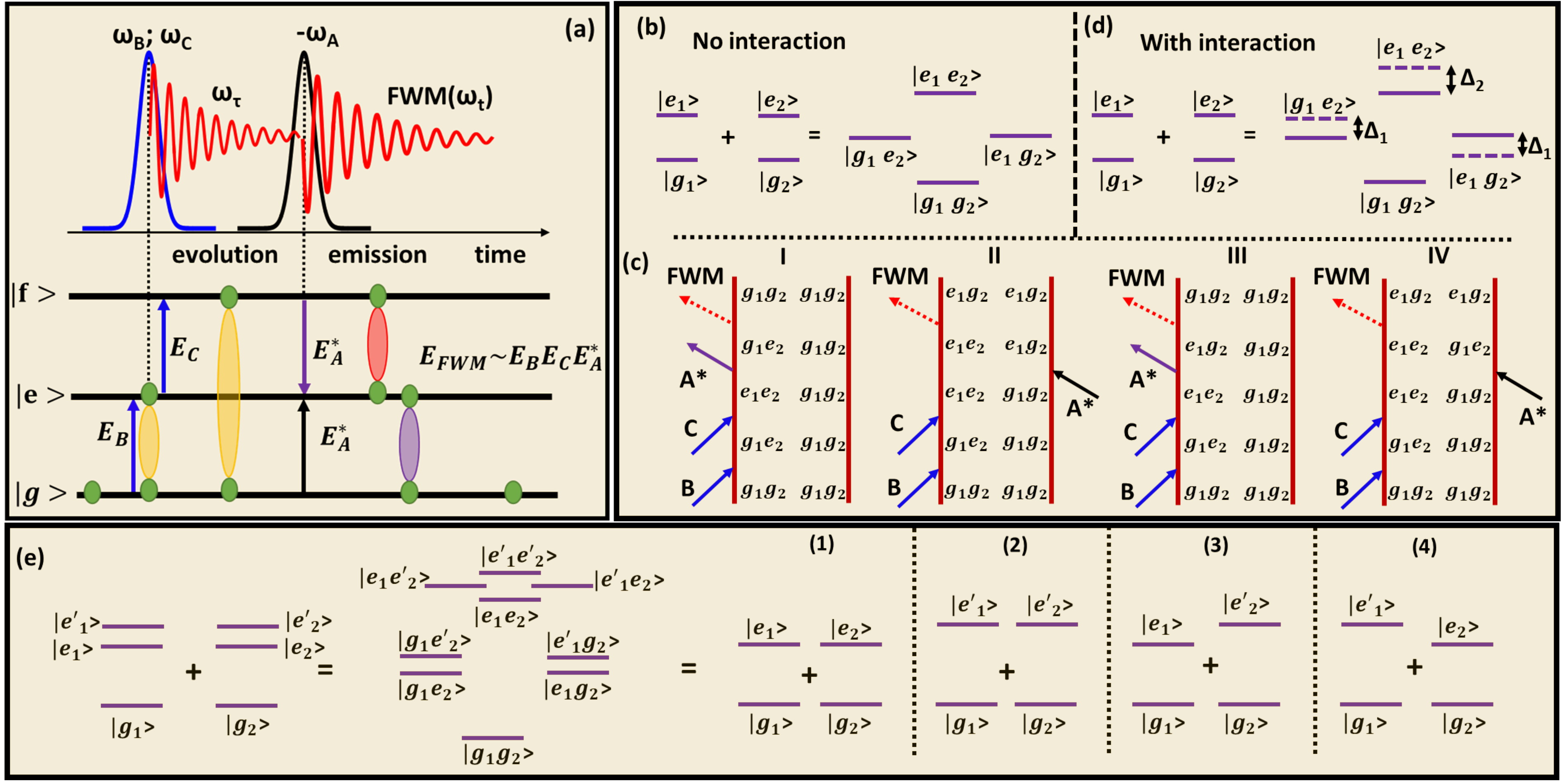}
\caption{\label{fig:DQ-MDCS} (a) Double-quantum excitation scheme. (b) energy level diagram of two combined two-level systems without interaction.  (c) Double-sided Feynman diagrams contributing for the generation of the FWM signal. (d)  energy level diagram of two combined two-level systems with interaction. $\Delta_{1}$ and $\Delta_{2}$ energy shifts due to interaction. Dotted lines show the shifted energy states. (e) Energy level diagram of two combined V-type systems which is also represented as a superposition of states created by two-level systems (1), (2), (3) and (4). $\ket{g}$, $\ket{e}$ and $\ket{f}$ correspond to ground, excited and doubly excited states, respectively.}
\end{figure*}

On the other hand if the complex phase-conjugated pulse arrives last (B, C, A*) then the corresponding 2D spectrum shown in Fig.~\ref{fig:MDCS_scheme}(c) (also referred to as a double-quantum 2D spectrum) can probe weak many-body interactions, for example long-range dipole-dipole interactions. \cite{Gao:16,Yang:mukame1,Yang:mukamel2,LomsadzePRL, Tollerud:16}. Pictorially the generation of a FWM signal in a simple three-level system is shown in Fig.~\ref{fig:DQ-MDCS} (a). The first pulse excites a coherence between the ground and singly excited states and then the same pulse (or different pulse if three pulses are used) excites the coherence between the ground state and the doubly excited state (also referred to as a double-quantum coherence). The last pulse then puts the system either back into the coherence between the ground and the singly excited states or into the coherence between the singly excited and doubly excited states. The final coherence then radiates the FWM signal that is detected. From this diagram it is clear that double-quantum MDCS is used to probe doubly excited states in the sample, however it can provide even more important information if applied to samples that do not have double excited states (or they are outside the bandwidth of the laser pulses). In that case the generation of a FWM signal can be described by combining two 2-level systems as shown in Fig.~\ref{fig:DQ-MDCS} (b) which clearly shows a doubly excited state. In Fig.~\ref{fig:DQ-MDCS} (c) we show the double-sided Feynman diagrams of the pathways that are contributing in the generation of the FWM signal. However, if there is no interaction between these two 2-level systems then the contributions have the same emission frequency, the same strength and the opposite sign (I-IV positive, II-III negative) and hence they cancel each other. The picture changes if we include many-body interactions (for example long range dipole-dipole interactions between two 2-level atoms). In this case, singly and doubly excited states experience slight energy shifts or changes in the line-width (Fig.~\ref{fig:DQ-MDCS} (d)). The changes break the symmetry between the contributions in Fig.~\ref{fig:DQ-MDCS} (c) which leads to the generation of a FWM signal. We note that the FWM signal is only due to the interactions (even a small interaction strength is enough to break the symmetry between the contributions) and hence double-quantum MDCS is a very sensitive and powerful tool for probing many-body interactions. We also note that experimentally, the separation of the FWM and linear signals are performed either utilizing a "Box" geometry configuration \cite{bristow} or a co-linear geometry and phase-cycling schemes \cite{tekavech, LomLet2017}.

Similar to single-quantum two-dimensional spectra, the peak lineshapes on double-quantum spectra can also provide critical information for example the underlying physics of the many-body interactions. However, the literature is not consistent about peak lineshapes (elongation) on double-quantum coherent spectra. In double-quantum 2D spectra the elongation of the peaks along the diagonal (for example the one shown in Fig.~\ref{fig:MDCS_scheme}(c)) suggests that there is correlation between excitation and emission frequencies. However, theoretically it has been shown that peaks are expected to be elongated and no correlation parameter was included in the calculation \cite{Tollerud:16}. Furthermore, there were several 2D experiments performed both on semiconductor materials and atomic samples that didn't show any peak elongation but instead they were vertically tilted \cite{Gao:16, Dai, Eric, Tollerud:16, Gael}. There were also experiments performed on atomic and molecular samples where the elongated peaks were observed \cite{LomsadzePRL, muka_elongation}. 

In \cite{LomsadzePRL} we briefly explained the experimental results using a simple model. In this paper we show the full theoretical model, which could also shed light on the inconsistencies between the results mentioned above. This simple model will help interpret experimental 2D spectra. In the next section we will show our model and the results of our simulation and in section III we will conclude our observation.

\section{Simulation and results}
To study the peak behavior of double-quantum 2D spectra we solved the optical Bloch equations for two coupled 3-level V-type systems. The energy level diagram of two combined V-type systems (without interaction) is shown in Fig.~\ref{fig:DQ-MDCS} (e) which can also be described as a superposition of the states created by coupled 2-level systems shown in Fig.~\ref{fig:DQ-MDCS} (e, 1,2,3,4). In our simulation we used infinitely short-pulses (delta-function pulses) $E(t)\approx E_{0}e^{i\omega t} \delta(t)$ and all of the excitation pulses were co-polarized. At first we treated the systems to be homogeneously broadened. Under these conditions the third order polarization (one of the pathways) for figure Fig.~\ref{fig:DQ-MDCS} (e(1))  created by the sequence of MDCS pulses (Fig.~\ref{fig:DQ-MDCS} (a)) is:  
\begin{equation}
    P(t,\tau)= (-i/\hbar)^3 E_{0}^{3}\mu_{ge}^4 exp[{i\omega_t t-i\omega_\tau\tau-\gamma_t t-\gamma_\tau \tau}]
\end{equation}
where $E_{0}$ is the magnitude of the excitation pulses (assumed to be the same for all three pulses), $\mu_{ge}$  is the transition dipole strength, $\omega_\tau$ is a double-quantum frequency, $\omega_\tau=\omega_t+\omega_A$
where $\omega_t$ is the emission frequency and  $\omega_A$ is the resonant frequency of the phase conjugated pulse.  $\gamma_t$ and $\gamma_\tau$ are the dephasing rates of the single and double quantum coherences, respectively. We note that in the model  $\gamma_t$ and $\gamma_\tau$ describe overall dephasing rates and we do not distinguish dephasing rates caused by spontaneous decay, power broadening, collision broadening etc.  In our model  $\gamma_\tau$=2$\gamma_t$  but one can model many-body interactions by including additional de-phasing rates that distinguishes single and double quantum coherences \cite{Tollerud:16}. In our simulation we modeled the interactions between the systems by including the energy shifts $\Delta_1$ and $\Delta_2$ (described below) for the single and double excited states (Fig.~\ref{fig:DQ-MDCS} (d)). 
 
To model a real system, inhomogeneous broadening was incorporated into the simulation by integrating the polarization over a generalised two-dimensional Gaussian function \cite{Steve:corre}:
\begin{widetext}

\begin{equation}
    f(x,y) = \frac{1}{2\pi\sigma_x\sigma_y\sqrt{1-\rho^2}}e^{-\frac{(\frac{x-\nu_x}{\sigma_x})^2-2\rho(\frac{x-\nu_x}{\sigma_x})(\frac{y-\nu_y}{\sigma_y})+(\frac{y-\nu_y}{\sigma_y})^2}{2(1-\rho^2)}}
\end{equation}

\end{widetext}
here $\nu_x$, $\nu_y$, $\sigma_x$, $\sigma_y$ correspond to the centers and widths of two interacting inhomogenously  broadened  resonances and $\rho$ is a correlation parameter. $\rho$=1, $\rho$=0, and $\rho$=-1  implies that the resonances are perfectly correlated, uncorrelated and anti-correlated, respectively.  

The integration yields
\begin{widetext}
\begin{equation}
\begin{aligned}
P(t,\tau)= (-i/\hbar)^3 E_{0}^{3} \mu_{ge}^4 ~ exp[{i\omega_{t}t-i\omega_\tau\tau-\gamma_tt-2\gamma_t\tau} -\frac{1}{2}(\tau^2(\sigma_A^2+\sigma_t^2+2\rho\sigma_A\sigma_t)\\
-2t\tau(\rho\sigma_A\sigma_t+\sigma_t^2)+t^2\sigma_t^2)] \end{aligned}
\end{equation}
\end{widetext}
If we assume that $\sigma_A$=$\sigma_t$ $\equiv\sigma$ and include the energy shifts due to interactions $\omega_t=\omega_{ge} \pm \Delta_1$ and $\omega_\tau=2\omega_{ge}+\Delta_2$ (where $\omega_{ge}$ is the transition frequency between the ground and the excited states), then all the polarization terms that are contributing in the generation of the FWM signal for the system shown in Fig.~\ref{fig:DQ-MDCS} (e(1)) are
\begin{widetext}

\begin{equation}
\begin{aligned}
P_I(t,\tau)= (-i/\hbar)^3 E_{0}^{3} \mu_{ge}^4 ~ exp[{i(\omega_{ge}+\Delta_1)t-i(2\omega_{ge}+\Delta_{2})\tau-\gamma_tt-2\gamma_t\tau}\\ -\frac{1}{2}(2\tau^2\sigma^2(1+\rho)-2t\tau\sigma^2(\rho+1)+t^2\sigma^2)] \\
P_{II}(t,\tau)= -(-i/\hbar)^3 E_{0}^{3} \mu_{ge}^4 ~ exp[{i(\omega_{ge}+\Delta_1+\Delta_2)t-i(2\omega_{ge}+\Delta_{2})\tau-\gamma_tt-2\gamma_t\tau}\\ -\frac{1}{2}(2\tau^2\sigma^2(1+\rho)-2t\tau\sigma^2(\rho+1)+t^2\sigma^2)] \\
P_{III}(t,\tau)= (-i/\hbar)^3 E_{0}^{3} \mu_{ge}^4 ~ exp[{i(\omega_{ge}-\Delta_1)t-i(2\omega_{ge}+\Delta_{2})\tau-\gamma_tt-2\gamma_t\tau}\\ -\frac{1}{2}(2\tau^2\sigma^2(1+\rho)-2t\tau\sigma^2(\rho+1)+t^2\sigma^2)] \\
P_{IV}(t,\tau)= -(-i/\hbar)^3 E_{0}^{3} \mu_{ge}^4 ~ exp[{i(\omega_{ge}+\Delta_2-\Delta_1)t-i(2\omega_{ge}+\Delta_{2})\tau-\gamma_tt-2\gamma_t\tau}\\ -\frac{1}{2}(2\tau^2\sigma^2(1+\rho)-2t\tau\sigma^2(\rho+1)+t^2\sigma^2)] \\
\end{aligned}
\end{equation}
\end{widetext}

Clearly without interactions the polarization terms cancel each other. Similar equations can be obtained for the systems shown in Fig.~\ref{fig:DQ-MDCS} (e-2,3,4) as well. 
A two dimensional spectrum is then generated by summing all the polarization terms and taking a two-dimensional Fourier transform with respect to $t$ and $\tau$. In our calculation we used $\sigma$=600 MHz (corresponding to Doppler Broadened atomic samples) and $\gamma_t$=6 MHz.

\begin{figure*}
\includegraphics[width=1\textwidth]{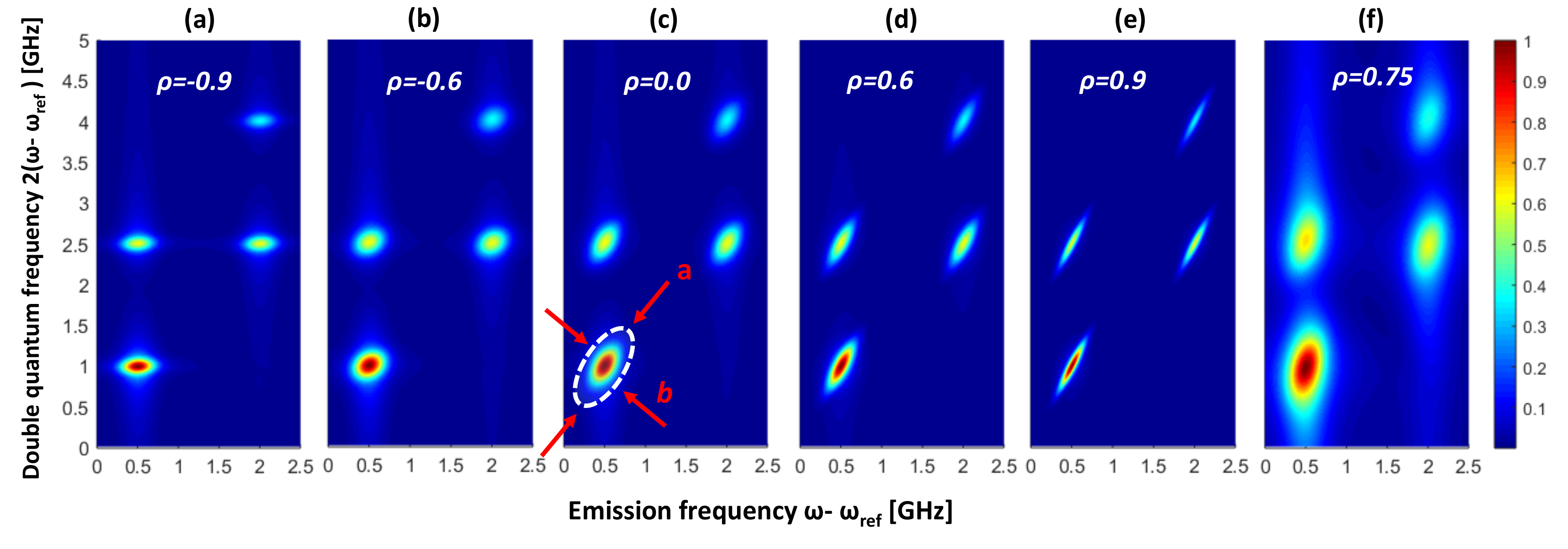}
\caption{\label{fig:Results} Simulation results. (a) $\rho$=-0.9 (b) $\rho$=-0.6, (c) $\rho$=0.0, (d) $\rho$=0.6, (e) $\rho$=0.9, (f) $\rho$=0.75 and increased decay rate.  $\omega_{ref}$ arbitrary optical frequency. Color scale shows normalized signal magnitude.}
\end{figure*}

In Fig.~\ref{fig:Results} we show the results. $\rho=0.0$  corresponds to un-correlated systems, $\rho=0.9$ - near perfectly correlated systems, $\rho=-0.9$ - near perfectly anti-correlated systems, respectively. $\rho=-0.6$ and $\rho=0.6$  correspond to partially anti-correlated and correlated systems, respectively. The peaks in each figure are diagonally elongated  and the effects of correlation is obvious. To give quantitative information, we measured the ellipticity of the peaks

\begin{equation}
    E = \frac{a^{2}-b^{2}}{a^{2}+b^{2}}
\end{equation}

where $a$  and $b$ are the sizes of the ellipse along the major and minor axes, shown in the figure. The measurements showed that for $\rho=0.0$ the ellipticity is 0.5 and it approaches to 0 and 1 for for $\rho=-1.0$ and $\rho=1.0$, respectively. It is important to note that the correlation parameter gives insight into the many-body interactions. For example, for Doppler broadened atomic systems near-perfect correlation implies that the generated FWM is due to the coupling of resonances between two atoms that have near zero relative velocity. On the other hand $\rho=0.0$ and  $\rho=-1.0$ correspond to coupling of the resonances of the atoms that have any relative velocity and opposite velocity, respectively. Experimentally a high degree of correlation (elongated peaks) $\rho=0.75$ has been observed in \cite{LomsadzePRL}. One can understand the results by comparing it to the runners in a sprint relay where two runners have the highest chance of transferring the baton if they have similar velocities. 

We also note there has been MDCS experiments performed on Doppler broadened atomic samples that showed that the peaks were not elongated along the diagonal line (the elongation was obscured and the peaks were elongated more along the vertical line) \cite{Gao:16, Dai}. But in the experiments an argon (Ar) buffer gas was introduced into the gas cell to artificially broaden (collisional broadening) the resonances to match the spectrometer resolution. To model the case we increased the decay rates (by a factor of 20 whihc is similar to the values of their experimental parameters) in our simulation. The results that are plotted in Fig.~\ref{fig:Results}(f) show that even with the high degree of correlation ($\rho=0.75$), peaks now are elongated along the vertical line which is similar to the results observed in \cite{Gao:16, Dai}.     

Our model can be extended to semiconductor materials as well. For quantum wells (and quantum dots) $\rho$ is expected to be close to zero (or partially correlated). This is because in semiconductor materials a double-quantum FWM signal is due to the coupling of the excitons that are located in nearby quantum wells and the thickness of wells are most likely random. In this case the peaks are expected to be elongated along the diagonal (ellipticity = 0.5) but the experiments showed that the peaks are tilted toward the vertical axis \cite{Eric, Gael}. This can be explained with the fact that unlike atomic systems, the excitons experience additional dephasing due to exction-exciton and exciton- free carrier scattering (which is a strong function of the temperature) reported in \cite{Tollerud:16}. This scattering causes the 2D peaks to be tilted similarly to the results that we showed for atomic systems in Fig.~\ref{fig:Results}(f).

\section{Conclusion}

In this work we theoretically investigated the lineshapes of double-quantum two-dimensional coherent spectra. We studied two coupled V-type systems and simulated the spectra by solving the optical Bloch equations. We showed that peak lineshapes describe how the resonances (from each system) are correlated and give insight into the mechanism of many-body interactions. We applied our model to Doppler-broadened atomic samples and explained the discrepancies between the experimental results reported in \cite{LomsadzePRL, Gao:16, Dai}.  We also discussed the expected peak lineshapes (and correlation) of 2D spectra generated in semiconductor materials and explained the difference between our theoretical results and the experimental results reported in \cite{Eric, Tollerud:16}.   
We hope that the MDCS community will benefit with this simple method when interpreting double-quantum correlation coherent spectra. 
We also hope that the model  will help the MDCS community understand the many-body interactions better, particularly the dipole-dipole interaction that plays the crucial role for photosynthesis and the formation of complex molecules.


\bibliography{References}

\end{document}